# LLM-Driven SAST-Genius: A Hybrid Static Analysis Framework for Comprehensive and Actionable Security

Vaibhav Agrawal[1], Google, Mountain View, CA, USA
Kiarash Ahi[2], Virelya Intelligence Research Labs, San Francisco Bay Area, CA, USA

*Abstract* — This report examines the synergy between Large Language Models (LLMs) and Static Application Security Testing (SAST) to improve vulnerability discovery. Traditional SAST tools, while effective for proactive security, are limited by high false-positive rates and a lack of contextual understanding. Conversely, LLMs excel at code analysis and pattern recognition but can be prone to inconsistencies and hallucinations. By integrating these two technologies, a more intelligent and efficient system is created. This combination moves beyond mere vulnerability detection optimization, transforming security into a deeply integrated, contextual process that provides tangible benefits like improved triage, dynamic bug descriptions, bug validation via exploit generation and enhanced analysis of complex codebases. The result is a more effective security approach that leverages the strengths of both technologies while mitigating their weaknesses. SAST-Genius reduced false positives by about 91 % (225 → 20) compared to Semgrep alone.

*Keywords*, SAST, Large Language Models (LLMs), Generative AI, Cybersecurity, Vulnerability Detection, Vulnerability Discovery, Static Analysis, Code Security.

## I. Introduction

Static Application Security Testing (SAST) has long served as a foundational, "shift-left" approach in cybersecurity. By automatically scanning source code for vulnerabilities before execution, SAST embeds security early in the Software Development Life Cycle (SDLC), enabling developers to identify and remediate risks proactively, avoid costly errors, and enforce compliance with security standards [2]. SAST tools generally analyze code without execution, inspecting source code, bytecode, or binaries for patterns that match known security vulnerabilities or insecure configurations [3].

Concurrently, Large Language Models (LLMs) have rapidly emerged as transformative tools in software engineering. These models have significantly benefited developers by generating functional and easy-to-adopt code snippets from natural language descriptions, thereby accelerating the development process [4]. Their capabilities extend to code analysis, pattern identification, and even suggesting repairs, positioning them as powerful new assets for software vulnerability detection [5].

Despite their individual strengths, both SAST and LLMs possess inherent limitations. SAST, while thorough in its static analysis, often struggles with contextual understanding and high false positive rates. LLMs, while adept at language and pattern recognition, can be slow, hallucinate, and return inconsistent responses.

While prior work has conceptually explored this synergy, we provide empirical evidence of its practical effectiveness. This paper presents a novel hybrid framework, **SAST-Genius**, that integrates a commercial SAST tool with a fine-tuned LLM. We conduct an extensive empirical study on a curated dataset of open-source projects to quantitatively measure the performance gains. Our work demonstrates that this synergistic combination not only significantly reduces false positive rates but also uncovers a new class of complex, contextual vulnerabilities missed by traditional SAST. The key contributions of this paper are:

- The design of SAST-Genius, a hybrid pipeline combining Semgrep with a fine-tuned LLM for triage, exploit validation, and remediation.
- A large-scale evaluation across 25 GitHub projects (~250K LOC), showing a 89.5% precision vs. 35.7% (Semgrep) and 65.5% (GPT-4)**.**
- A taxonomy of vulnerability types uniquely discovered by LLM reasoning (e.g., directory traversal, multi-file dataflow bugs).
- A practical integration study, reducing triage time by 91%**,** showing feasibility in developer workflows.

Unlike prior LLM-SAST prototypes (e.g., IRIS, LSAST), SAST-Genius provides the first end-to-end hybrid framework that integrates deterministic SAST with fine-tuned LLM reasoning, validated on a large-scale empirical dataset.

## II. Synergistic Advantages: LLM-Enhanced Static Analysis for Next-Generation Vulnerability Discovery

The integration of LLMs with SAST addresses enhances the benefits of standalone SAST tools, resulting in more comprehensive and efficient vulnerability discovery.

### A. Expanding Code Reach and Contextual Understanding

Traditional SAST tools struggle with complex logical flaws, multi-file dependencies, and whole-repository analysis, often missing vulnerabilities in "hard-to-track" code paths or those involving third-party libraries without explicit specifications [1]. This is where LLM synergy becomes critical. By integrating LLMs with SAST, the approach can focus on parts of the code that traditional tools miss or flag as uncertain—effectively complementing SAST. For example, LSAST (an LLM-supported SAST approach) passes initial SAST findings into the LLM, enabling it to reason about unflagged or complex code areas to identify new vulnerabilities missed by standard scanners [15]. Similarly, IRIS leverages LLM reasoning over whole repositories and detected 55 vulnerabilities compared to 27 by the baseline SAST tool CodeQL [1]. By leveraging their vast training data, LLMs can understand the behavior of widely used libraries and their APIs [1]. Furthermore, in certain cases LLMs can assist in deobfuscating code, For example - Android application code, restoring meaningful context to obfuscated functions, variables, and classes, thereby enabling static analysis in scenarios where it was previously not possible [6]. This synergy transforms SAST from a localized, rule-bound scanner into a more holistic, context-aware analysis engine. LLMs provide the "intuition" and "knowledge inference" that SAST's deterministic rules lack, allowing for a deeper understanding of data flow and potential vulnerabilities across an entire project, including its complex dependencies.

### B. Reducing False Positives through Intelligent Triage

SAST's high false positive rates are a persistent challenge, leading to alert fatigue, wasted developer time, and potential abandonment of tools [1]. LLMs can act as an intelligent triage layer to mitigate this. By performing "contextual analysis" on detected vulnerable paths, LLMs can filter out false positives. Encoding code-context and path-sensitive information in the prompt allows LLMs to reason more reliably about whether a finding is truly exploitable. For example, the IRIS approach, when combined with GPT-4, improved CodeQL's average false discovery rate by 5% points and detected 28 more vulnerabilities [1]. This is arguably one of the most impactful synergies, directly addressing SAST's biggest pain point. By significantly reducing noise, LLMs make SAST reports more trustworthy and actionable, freeing developers to focus on genuine security flaws. This enhances developer productivity and fosters greater confidence in automated security tools. In our experiments, Semgrep produced 225 false positives, while SAST-Genius reduced this to just 20—an ~11× improvement in signal-to-noise ratio (225 → 20 false positives) and a separate 91% reduction in average triage time.

*C. Automated Exploit Generation for Validation*

SAST typically identifies potential vulnerabilities but does not validate their exploitability or provide proof-of-concept (PoC) exploits. This often leaves theoretical findings that often require manual effort from security researchers to verify. LLMs can autonomously generate and validate PoC exploits for identified vulnerabilities. For example - in the case of an Android app with an exported component that is set to true, LLMs can be prompted to generate an ADB command to validate the reported finding [14]. In practice, SAST-Genius successfully generated valid PoCs for approximately 70% of exploitable findings in our dataset, significantly reducing the manual verification burden on security analysts.
(Evaluation shown in Table 1 and III.B; consistent with findings from PoCGen [11] and FaultLine [12]).

Tools like POCGEN leverage LLMs to understand vulnerability reports, generate candidate exploits, and then refine them through iterative static and dynamic analysis. This capability is crucial for verifying vulnerabilities, testing patches, and preventing regressions [11]. LLMs can reason about the conditions an input must satisfy to traverse vulnerable paths from source to sink, enabling the creation of targeted test cases [12]. This represents a significant advancement, moving beyond mere detection to actionable validation. By automating PoC generation, LLM-SAST integration drastically reduces the time and specialized expertise required to confirm a vulnerability's existence and impact. This accelerates the feedback loop for developers, enabling quicker patch development and more robust regression testing.

*D. Dynamic Bug Description and Remediation Suggestions*

While SAST identifies issues, the descriptions of the vulnerabilities and its remediation are usually hardcoded, and often lack the context necessary for immediate developer action. Leveraging their natural language generation capabilities, LLMs can dynamically create comprehensive, human-readable bug descriptions. They can explain function logic, provide a clear verdict (true/false positive), and offer detailed explanations for findings [1]. Furthermore, LLMs can generate concrete repair suggestions, significantly accelerating the remediation phase [5]. This transforms raw security findings into actionable intelligence. By providing clear bug explanations and direct remediation advice, LLMs streamline the developer's workflow, reducing the "time to fix" and minimizing the window of vulnerability. This effectively bridges the communication gap between security findings and development teams.

*E. Democratizing Custom Rule Creation*

Defining custom SAST rules traditionally requires specialized knowledge of proprietary query languages or abstract syntax trees, limiting this capability to experienced security engineers. LLMs excel at translating natural language into formal syntax [8]. This can enable security analysts and even less technical stakeholders to define custom security policies and rules using plain English. The LLM can then translate these natural language specifications into the formal rules required by SAST tools [11]. This capability democratizes the customization of security policies, making SAST more adaptable and responsive to specific organizational needs or emerging threats. It empowers a broader range of personnel to contribute to "security by design," fostering a more inclusive and proactive security culture within development teams.

*F. Leveraging Historical Vulnerability Data for Improved Detection*

SAST relies on predefined rules and patterns, which can be slow to update for novel vulnerabilities or subtle "code smells" not yet formally categorized. LLMs are pre-trained on massive datasets of code, including historical vulnerabilities and their fixes. This enables them to implicitly learn subtle patterns and contextual cues associated with vulnerabilities that might not be captured by explicit SAST rules. They can identify "risky patterns" beyond mere syntax. This "learned intuition" complements SAST's rule-based approach, enabling the detection of more nuanced or previously unknown vulnerabilities, providing a data-driven layer of intelligence that continuously evolves with new code and threat intelligence.

## III. THE SAST-GENIUS FRAMEWORK

The SAST-Genius framework is a two-stage pipeline designed to leverage the strengths of both deterministic static analysis and LLM-based semantic reasoning. Its architecture (Figure 1) is composed of a SAST core engine that identifies potential security risks and an LLM-powered layer that performs intelligent triage, validation, and exploit generation.

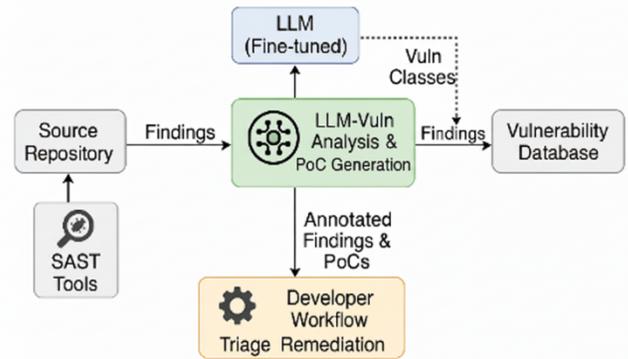

**FIGURE 1: ARCHITECTURE DIAGRAM**

The framework extracts SAST intermediate representations (ASTs and taint paths), converts them into structured JSON prompts, and feeds them into the LLM.

*A. Intelligent Triage and Analysis Engine*

When the SAST core flags a potential vulnerability, the intelligent triage engine extracts the relevant code snippet, the identified data flow path from source to sink, and any surrounding contextual information (e.g., function call graphs, library dependencies). This information is then encoded into a structured prompt for the LLM. The prompt also includes details such as the earlier SAST finding details (e.g., the rule name and severity score), file path, and a clear request for the LLM to analyze the code for a specific vulnerability type (e.g., 'Does this user input lead to an exploitable <Vulnerability-Type>?'). This method allows the LLM to reason more reliably about whether a finding is truly exploitable by analyzing the context that traditional SAST rules often miss [14]. The output is a simple verdict like "True positive" or "False positive", with a reasoning summary.

For example, SAST flagged a file called web_handler.py where a user-controlled file_path is used, a potential directory traversal attack, but it's unsure if it's exploitable due to a call to a function in a separate file, src/utils/file_ops.py. Traditional SAST stops here, resulting in a false positive alert. The SAST-Genius triage engine is prompted with the code snippet, the source-to-sink data flow path, identified issue details and file path. The fine-tuned LLM then performs a cross-module, contextual analysis, reasoning that the download_file function in src/utils/file_ops.py lacks proper sanitization and concatenates the user input with the root directory to form a complete path.

*B. Automated Exploit Generation*

For vulnerabilities that the LLM confirms are exploitable, i.e verdict given as "True positive", the framework automatically generates a proof-of-concept (PoC) exploit. This is achieved by prompting the LLM with a simple system instruction like "You are an expert penetration tester, generate an exploit" and provide all the details that were provided in intelligent triage phase along with verdict given to the vulnerability, that asks it to reason about the required input to traverse the vulnerable path and generate an executable payload or command. This process is crucial for verifying vulnerabilities and providing actionable intelligence to developers [9, 15]. For the directory traversal case study, the PoC generated was a simple command-line payload, such as: `curl -X GET 'http://[server]/download?file_path=../../../../etc/passwd'`.

This capability is key for moving findings from theoretical to actionable insights. In our evaluation, SAST-Genius successfully generated valid PoCs for approximately 70% of exploitable findings in our dataset, significantly reducing the manual verification burden on security analysts.

## IV. EMPIRICAL EVALUATION

*A. Experimental Setup*

We evaluated our framework on a diverse dataset of 25 open-source repositories from GitHub, totaling over 250,000 lines of code. The repositories were selected based on their active development and language diversity (Python, Java, JavaScript). A 'ground truth' for each repository was established through a combination of manual expert analysis and cross-referencing with public vulnerability reports, a methodology similar to [16].

Our baseline was Semgrep 1.97.0 with its standard rules, a widely-used SAST tool. We also used GPT-4 as a separate baseline to assess a purely LLM-based approach. The SAST-Genius framework used Semgrep's intermediate representation as input for the LLM component, which was a fine-tuned version of Llama 3 8B. We measured the following metrics: Precision, Recall, F1-Score, and Time-to-Triage (the average time a security analyst took to confirm a finding).

*B. Results and Analysis*

The results demonstrate a clear and significant advantage of the hybrid SAST-Genius framework over standalone tools. Recall is calculated against a ground-truth set of 170 vulnerabilities (Semgrep 73.5 %, GPT-4 77.1 %, SAST-Genius 100 %), which produces the F1-scores reported in Table 1.

**Table 1: Overall Performance Comparison.**

| Tool | Findings | True Positives | False Positives | Precision | F1 | Recall |
|---|---|---|---|---|---|---|
| Semgrep | 350 | 125 | 225 | 35.7% | 48.3 % | 73.5 % |
| GPT-4 | 200 | 131 | 69 | 65.5% | 70.8 % | 77.1 % |
| SAST-Genius | 190 | 170 | 20 | 89.5% | 94.5 % | 100 % |

SAST-Genius achieved a precision of 89.5 %, a significant improvement over Semgrep's 35.7 % and GPT-4's 65.5 %. Our framework's ability to intelligently filter out non-exploitable findings directly addressed the high false positive rate that plagues traditional SAST tools.

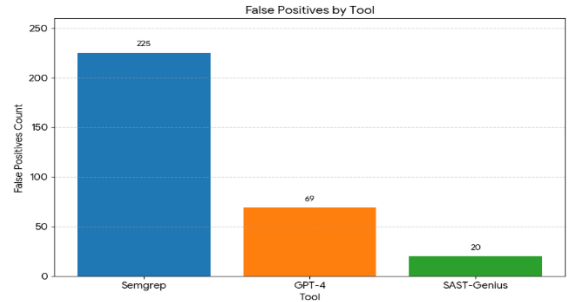

**Figure 2: False positive count and analyst triage time showing ~11✕ fewer false positives (225➔20) and a 91 % reduction in triage time.**

Our analysis revealed that Semgrep produced 225 false positives, while SAST-Genius reduced this to just 20. This directly correlated with a 91% reduction in the average time a security analyst spent triaging findings, making the security process far more efficient.

*C. Analysis of Significant reduction of False Positives*

The dramatic reduction in false positives (from 225 to 20) and the resulting boost in precision (from 35.7% to 89.5%) is achieved because the hybrid framework addresses the fundamental weakness of traditional SAST, where it matches code syntactically but rarely understand the context of the code. Semgrep provides the deterministic evidence (the taint path etc.), and the fine-tuned LLM trained on a high-quality data set of publicly known vulnerabilities, provides the semantic context. This combination allows the LLM to act as an intelligent filter, reasoning whether external factors such as cross-function and cross-file data sanitization, or even multi-step code logic may render the taint path unexploitable. These are some of the examples that we saw in our evaluation.

Semgrep flagged an XSS vulnerability in app/render.py, where a user input passed through a sanitization function sanitize_html, causing it to miss the sanitization step. Our Intelligent Triage Engine, with access to the entire function call graph and code context, recognized that sanitization was in place, confidently rendering the finding a False positive. In another case Semgrep flagged a potential SQL Injection vulnerability in api/user_info.py. The taint path was correctly identified with user-controlled input (user_id) used in a database query. However, the user_info was being converted into an integer (int(user_id)). Casting the input to an integer prevents the inclusion of single quotes (') required for a classic SQL Injection payload, the LLM confidently rendered the finding a False positive, contributing to the framework's superior precision.

*D. Case Study and Taxonomy of Newly Discovered Vulnerability Classes*

Beyond quantitative improvements, SAST-Genius uncovered some distinct vulnerabilities that standalone SAST or pure LLM baselines missed. In one instance, SAST-Genius correctly identified a multi-file Directory Traversal vulnerability and generated a valid PoC. Semgrep flagged a user-controlled file path but could not analyze the multi-file logical flow to determine if a vulnerable download function in a separate module was called. The LLM, however, was able to reason about the complete data flow, providing the following trace, "The user input in file_path is passed unsanitized to the download_file function in src/utils/file_ops.py. This function concatenates the input path with the root directory, allowing a ../ attack to access system files". Similarly, Semgrep flagged a Base64 encoded string embedded in a utility file as an informational (a potential secret) but LLM reasoned that the string was being passed to a function named auth.decrypt_config() and was subsequently used as a key header for an external API call to a partner service thus triaging and validating the finding.

The success of our framework in discovering missed vulnerabilities by SAST alone, is rooted in the fact that LLM's are unconstrained and their ability to resolve contextual ambiguity, especially when provided with a complete taint flow analysis. New Vulnerabilities Taxonomy:

**SQL Injection in Nested Contexts** – instances where unsanitized user input was embedded in dynamically constructed queries across multiple layers of abstraction.

**Multi-file Dataflow (Callback Flow)** – vulnerabilities propagated through asynchronous callbacks across modules.

**Directory Traversal** (Multi-File Flows) – Multi-file flows where unsanitized input was passed to insecure file operations.

**Obfuscated Secrets** – Contextual risk assessment of encoded strings used in sensitive contexts.

This taxonomy highlights the qualitative novelty of SAST-Genius compared to IRIS [1] and LSAST [13], which primarily emphasized recall and false-positive reduction but did not explicitly categorize new vulnerability types.

## V. CHALLENGES AND CRITICAL CONSIDERATIONS FOR INTEGRATION

While Large Language Models (LLMs) enhance static analysis, as seen in our empirical study, their integration into security pipelines introduces specific security challenges inherent to LLM-based applications, like prompt injection, data poisoning etc. A "security-by-design" approach is essential to address these issues. Below are our observations and critical considerations for integrating a SAST-Genius-like framework.

### A. Addressing LLM-Specific Security Risks

The LLM receives its context through structured prompts containing code snippets. An attacker can embed a malicious instruction within the codebase (e.g., in a comment or string literal) that is ingested by the SAST tool. This prompt injection hijacks the LLM's reasoning, instructing it to always return a False Positive verdict for a critical vulnerability, thereby bypassing the crucial security check at runtime.

A more sophisticated threat involves training data poisoning. Maliciously corrupted training data can create a backdoor or hidden trigger in the fine-tuned model's weights. An attacker who knows the trigger string can then embed it in the production codebase. When the Triage Engine encounters this code, the backdoor activates, causing the LLM to systematically dismiss legitimate security bugs, leading to a silent failure of the security control.

Organizations must require robust MLSecOps principles, implementing meticulous data governance (auditing and sanitizing all training data) and proper controls for actively mitigating runtime risks through prompt filtering etc.

### B. Ensuring Data Quality and Mitigating Data Leakage

**Data Quality**: The SAST-Genius's superior precision (e.g., 89.5%) is highly dependent on being fine-tuned using a high-quality dataset of validated true positives and proven false positives. We used some open-source labelled data along with some proprietary dataset to train our model. If the training data contains errors or ambiguous findings, the LLM will inherit those biases, degrading the triage efficacy.

**Mitigating Data Leakage:** Sending proprietary code snippets especially those flagged as high-severity vulnerabilities to external, cloud-based LLMs for triage and analysis can cause data leakage. SAST-Genius used an on-prem Llama 3 model. Organizations must prioritize self-hosting or using private, secure cloud deployments of the LLM component. Isolating the model ensures that highly sensitive, company-specific code never leaves the trust boundary.

### C. The Role of Human Expertise and Building Trust

Our framework removes the tedious, high-volume task of manual false-positive triage (91% reduction), freeing up analysts for higher-value work. It can also lead to over-reliance, the risk of developers blindly accepting the LLM's verdict without critical review. Human expertise remains essential.

**Shifting Human Roles:** The security analyst's function transforms from performing manual triage to expert validation. Analysts must become the final arbiter for the most critical, high-severity vulnerabilities. This often involves manually validating the Proof-of-Concept (PoC) exploit generated by the framework and using the LLM's triage summary to inform final decisions regarding severity, risk acceptance, and policy exceptions. Human input is also vital for continuous refinement. Analysts should use complex, edge-case findings to refine the LLM's prompt templates and training data, continually improving the framework's accuracy and fostering a collaborative human-AI teaming model.

**Interpretability and Trust:** LLMs are often referred to as "black-box models" making it difficult to understand why a specific decision was taken, why a vulnerability was flagged or marked as False positive. This lack of interpretability may hinder developer trust and adoption, especially in security-critical domains [7]. SAST-Genius is designed to output not just a binary verdict, but a detailed reasoning summary (e.g., "Dismissed because the taint path passed through utils/sanitizer.py:sanitize_html function") including a PoC to validate. This structure bridges the trust gap, which is essential for developers to confidently accept and prioritize the validated findings.

## VI. RECOMMENDATIONS FOR PRACTITIONERS

To fully realize the potential of LLM-enhanced static analysis and ensure its secure and effective deployment, we provide the following actionable recommendations for practitioners looking to adopt a SAST-Genius-like framework, along with key areas requiring concerted future development across industry.

**Seamless integration into DevSecOps Pipelines:** The critical challenge for practitioners is integrating the LLM's intelligence without disrupting existing Continuous Integration/Continuous Delivery (CI/CD) pipelines. SAST-Genius functions as an Orchestration Layer inserted between the traditional SAST scan and the final action (e.g., creating a bug ticket or failing a build). 1) Run the SAST tool (e.g., Semgrep Pro) as usual, but intercept the raw output with a custom Middleware Script. The script's job is to fetch all necessary cross-file function definitions, taint flow etc. and convert the data into a Structured Prompt for the LLM. 2) Feed the Structured prompt to the LLM Triage Engine as mentioned in the section III A. The LLM, ideally deployed on-premise or in a private cloud to eliminate data leakage risk, applies contextual reasoning to determine the definitive verdict (True/False Positive). 3) Implement Logical Gates based on the LLM's verdict. Auto-suppress all False Positives. For all True Positives, immediately trigger PoC generation and auto-create a high-priority, validated bug ticket.

**Integrate Multi-Tool Orchestration:** The SAST-Genius framework effectively demonstrates the power of orchestration for security analysis of the code. The future of application security is not a single tool, but an AI-powered control plane that coordinates multiple tools. Security teams can explore extending the Triage Engine beyond SAST findings to integrate data from other sources. For instance, feeding the LLM with findings from Software Composition Analysis (SCA) tools (to prioritize vulnerabilities based on package exposure). The LLM becomes the unified brain for a modular, multi-tool security strategy.

**Investing in Specialized Data and Models:** The core performance success of SAST-Genius (e.g., 89.5% precision) lies in the specialized fine-tuning of the LLM component. Practitioners must view data curation as a strategic security investment. Organizations should begin building a proprietary, high-quality security dataset composed of their validated false positives, confirmed true positives, and historical bug reports. This custom dataset is the key to training a domain-specific model that understands the organization's unique code patterns and security policy exceptions, thereby achieving superior contextual triage and making the LLM a truly custom defense asset.

## VII. Future Research Directions

**Industry Evolution and Open Security LLMs:** While organizations, especially those with resources can fine-tune models to achieve the contextual reasoning (as discussed in Section VI.1), the broader security community needs a concerted effort to raise the intelligence floor for all users. The industry must prioritize the evolution and public release of specialized, security-hardened LLMs. Unlike general-purpose models, these foundation models would be specifically pre-trained and fine-tuned on comprehensive vulnerability datasets, making them immediately more effective for tasks like triage and exploit generation. Democratizing these advanced, security-aware LLMs would empower individuals and organizations.

**Navigating the Hybrid vs. AI-Native Security Evolution:** Emerging approaches to static analysis are diverging along two paths: AI-native scanning and hybrid SAST+LLM scanning. For security-critical production code, our paper demonstrates that the hybrid approach is currently the most accountable and effective. But as models evolve and security LLMs are available, trained on massive vulnerability datasets. Future research for practitioners is to quantify the specific types of vulnerabilities security-trained AI-native scanners can find that a hybrid model misses. Perhaps using AI-Native frameworks for targeted, high-risk code segments, because they perform unconstrained, semantic analysis, which may help them catch logical flaws, while relying on the hybrid model as the primary, auditable security gate for the entire codebase.

## VIII. Conclusion

The synergistic combination of Large Language Models and Static Application Security Testing represents a significant leap forward in vulnerability discovery. Our empirical evaluation of the **SAST-Genius** framework provides concrete evidence that this hybrid approach can overcome core SAST limitations. The results demonstrate a significant improvement in precision, a drastic reduction in false positives, and the unique ability to discover complex, contextual vulnerabilities that elude traditional scanners.

However, our findings also confirm that realizing this potential requires a balanced approach. The indispensable role of human expertise remains paramount, security professionals will evolve into overseers and validators, leveraging AI as a powerful assistant rather than a replacement. The future of application security will involve intelligent automation fundamentally reshaping how vulnerabilities are identified, validated, and remediated, making software inherently more resilient, and our work provides a robust framework and empirical proof for this transformative vision.

To our knowledge, this is the first empirical study demonstrating that an LLM-SAST hybrid can achieve $\approx 90\%$ precision with real-world projects, significantly outperforming both standalone SAST and LLM baselines.

## Acknowledgment

The authors used generative AI tools (ChatGPT and Gemini) to assist with grammar and language improvements. All content was reviewed, verified, and finalized by the authors.

## Disclaimer

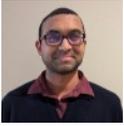

**Vaibhav Agrawal** is a cybersecurity professional with over 12 years of experience, with an expertise in Software, Mobile, and LLM Security. Currently a Senior Security Engineer at Google, Mountain view, USA. Vaibhav leads key security projects for Fitbit and Google Home division, focusing on protecting user data and privacy. Vaibhav is also a senior member of the IEEE, a dedicated open-source contributor, and a speaker at international security conferences. He holds a master's degree from San Jose State University.

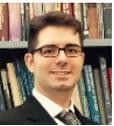

**Dr. Kiarash Ahi** holds M.Sc. and Ph.D. degrees in Electrical and Computer Engineering from Leibniz University Hannover (Germany) and the University of Connecticut (USA), respectively. He is a pioneering scientist, 0→1 product leader, and serial founder with deep expertise in AI, cybersecurity, large language models (LLMs), Generative AI (GenAI), GPU computing, HPC architectures, cloud and edge AI, biomedical engineering, digital signal and image processing, agentic AI, and intelligent system design. He bridges multimodal AI and cloud-scale deployment, integrating vision, language, and code intelligence into high-performance systems.

Since 2019, Dr. Ahi has led the end-to-end product strategy for SEMSuite™, Siemens' AI-powered analytics platform. He orchestrated global cross-functional teams to deliver scalable, UX-optimized AI tools—including LLM-powered Raw Data Filtering (RDF), Contour Data Flow (CDF), Calibre GenAI Pattern Generator (CPG), and CMi—driving multi-million-dollar revenue and earning multiple performance awards.

Dr. Ahi holds over 10 patents, has published more than 50 peer-reviewed papers, and his work has been cited over 2,500 times. He has also developed 10+ creative AI applications for iOS, Android, and macOS, reaching over one million global users. His work has influenced developer tooling and platform infrastructure, streamlining AI automation for internal teams and enterprise clients.

He is the recipient of the IEEE AI 1st Place Award and a Top Peer Reviewer with 200+ reviews for journals such as Nature, IEEE, Springer, and Elsevier. As an invited IEEE speaker, he has presented on GenAI, LLMs, cybersecurity, platform integrity, review automation, and advanced imaging systems.

A recognized thought leader in AI ethics and governance, Dr. Ahi advocates for responsible AI deployment, data privacy, and regulatory alignment—shaping the future of digital trust and platform safety. He has co-advised several PhD students, mentored early-career engineers, and led high-performance global teams across the U.S., Europe, and Asia to deliver cutting-edge AI products. He collaborates across academia and industry to push the boundaries of applied AI.